# CWTS crown indicator measures citation impact of a research group's publication oeuvre

Henk F. Moed

*Senior scientific advisor at Elsevier, Amsterdam, and former professor of research assessment methodologies at Leiden University, the Netherlands.*

The article "Caveats for the journal and field normalizations in the CWTS ("Leiden") evaluations of research performance", published by Tobias Opthof and Loet Leydesdorff (Opthof & Leydesdorff, 2010), denoted as O&L below, deals with a subject as important as the application of so called field normalized indicators of citation impact in the assessment of research performance of individual researchers and research groups. Field normalization aims to account for differences in citation practices across scientific-scholarly subject fields.

O&L claim that the subject field delimitation in the CWTS studies is questionable if not invalid as there are strong overlaps between fields while better alternatives are available. Their central claim is that the CWTS field normalized indicator (Moed, de Bruin & Van Leeuwen, 1995; Van Raan, 2004) is "seriously flawed because one divides averages instead of averaging divides". Finally, they underline the importance of transparency and traceability of indicators, perhaps suggesting that CWTS assessment methodology violates these principles.

As the primary author of the papers presenting the "Leiden" indicators and of many reports and articles reporting on the outcomes of assessments actually using these measures, I would like to comment in three separate sections on each of these three main issues O&L addressed.

## 1. Subject field delimitation

The subject classification used in the CWTS studies is a grouping of journals into subject categories developed at Thomson Reuters and adapted by CWTS. It is true that these journal categories are partly overlapping. But the use of overlapping categories lead to incorrect normalizations only if the citation characteristics in them are different from one another, but O&L do not show that this is actually the case.

But more importantly, I agree that alternative subject classification methodologies are feasible. A better approach to subfield delimitation from a bibliometric viewpoint is one that distinguishes specialist journals from more general sources covering an entire discipline or even science as a whole. In a first step one groups specialist journals into specialties. Next, one allocates papers in general journals to these specialties on a paper-by-paper basis, based on reference analysis, thus splitting up such general journals. In this way, for instance, astronomy papers in the multidisciplinary journal Nature would be allocated to a category covering specialist astronomy journals if they cite these specialist journals in their reference lists. I refer to the work of Lopez-Illescas et al. (2009) for an application of this approach to the delimitation of a medical specialty – oncology, and to Glanzel, Schubert & Czerwon (1999) for a method to split up multidisciplinary or general journals.

## 2. Globalized versus averaged ratios

O&L's key point is that the normalization procedure underlying the CWTS 'crown' indicator of field normalized citation impact is invalid. A field normalized indicator computed for a particular set of papers 'divides' in some way the actual citation rate of the papers in the set with the average citation rate of articles in the subject field(s) covered by these papers. The issue at stake is: how precisely is this 'division' or 'ratio' defined? By first calculating for each individual paper the ratio of its actual citation count and the subject field average, and next compute the average of this ratio over all papers in the set? Or by first calculating the average citation rate of all papers in the set, and next calculate the ratio of this average and the subject field average?

Egghe & Rousseau (1996) denoted these two types of ratios as 'averaged' and 'globalized', respectively. Adopting a mathematical-statistical viewpoint they claimed that in most applications the latter is better than the former. Other authors promoted averaged ratios (e.g., Rehn & Kronman, 2008) or used both types (SCIMAGO, 2009). O&L not only claim that averaged ratios are statistically better, but also that globalized ones are invalid. They even maintain that a globalized ratio violates a mathematical principle as severe as the '*Please Excuse My Dear Aunt Sally*' rule in algebra.

I would like to briefly explain why we have chosen to construct a globalized measure of citation impact and highlight its theoretical-conceptual background (Moed, 2005, pp.216-218). It follows from the view of research articles as elements from coherent publication ensembles of research groups carrying out a research programme. Citing authors acknowledging a research group's works do not distribute their citations evenly among all papers emerging from its programme, but rather cite particular papers that have become symbols or 'flags' of such a programme. Citations to these flag papers can be conceived as citations to the entire oeuvre and to the programme embodied in it. The way in which the citations to a group's oeuvre are distributed among the papers in the oeuvre is not relevant. If an oeuvre is cited in total, say, 100 times, it is not relevant whether there are two papers cited 50 times, or 10 papers cited 10 times each. In both cases the group's normalized citation impact should be the same. The CWTS, globalized, 'crown' indicator has this property, wheras the averaged variant has not.

In the calculation of a globalized measure citations are in a sense detached from the papers formally receiving them. The total number of citations to an oeuvre is compared to (divided by) the expected number of citations of a set of papers with the same size, and the same distribution across subject fields and, in the case of the CWTS crown indicator, across types of papers and publication years. This measure can be labeled as a group's field-normalized oeuvre impact, and the averaged version advocated by O&L as a group's average field normalized impact per paper.

Two additional comments should be made. Firstly, the theoretical considerations presented above explicitly speak of research groups publishing a cognitively coherent publication oeuvre. One can ask whether a globalized ratio is an appropriate measure to express the citation impact of aggregations of groups, such as entire universities. My reply is that I believe that it does not make much sense calculating one single index for an institution as diverse in subject coverage as a university (AUBR Expert Group, 2009). Computing indicators per subject field would be much more informative, and precisely at the level of a subject field the difference between the values of a globalized and an averaged measure vanishes.

My second comment is that I would strongly encourage conducting more research on the differences between globalized and averaged impact ratios at the level of research groups and other aggregations. O&L have rightly underlined the prominent position this topic deserves on the research agenda in our field. This research should focus not merely upon mathematical-statistical aspects, but also upon a further theoretical foundation of what citations measure and why citation distributions are skewed, and, last but not least, upon validation of the results against outcomes of peer assessments and other types of indicators.

## 3. Traceability and transparency

O&L advocate traceability and transparency of scientometric indicators used in the policy arena. Recognizing that other principles are essential as well, for instance, privacy rules – O&L respect these as they do not publish the names of the researchers for which they present indicator results –, I fully agree with O&L that traceability and transparency are important principles. The more frequently indicators are used, the more important these principles become. The reverse statement is true as well.

I want to highlight relevant facts related to these two aspects that O&L do not mention in their paper. The CWTS "bottom-up" methodology enables researchers under assessment to verify the bibliometric data collected about their oeuvres, or their commissioning organizations to deliver authorized publication lists. It provides final detailed outcomes of a group not only to the commissioning organization but also to the group itself. Detailed outcomes are embedded in a report underlining the potentialities and limitations of bibliometric indicators in general, and discussing factors that one should take into account when interpreting the outcomes.

The Leiden methodology is founded on the notion that the use of citation analysis in the assessment of individuals, groups and institutions is more appropriate the more it is formal – i.e., known to all that indicators are used as one of the sources of information; open – those subjected to the bibliometric analysis have the opportunity to examine the accuracy of underlying data, and to provide background information; scientific-scholarly founded; supplemented with expert and background knowledge; carried out in a clear policy context; stimulating users to explicitly state basic notions of scholarly quality; and used in a enlightening rather than formulaic manner, aimed at obtaining insight rather than being used as inputs in funding or rating formulas.

## Acknowledgement

The author is grateful to Martijn Visser and Thed van Leeuwen at the Centre for Science at Technology Studies (CWTS) for their comments on a draft version, and for their support of the ideas outlined in this letter.